\documentclass[onecolumn,aps,showpacs,prb,preprint,superscriptaddress]{revtex4}
\usepackage{tipa}
\usepackage{bbding}
\usepackage{txfonts}
\usepackage{amssymb}
\usepackage{graphicx}
\usepackage{CJK}

\begin{document}

\title{Modulation of the thermodynamic, kinetic and magnetic properties of the hydrogen monomer on graphene by charge doping}

\author{Liang Feng Huang}\affiliation{Key
Laboratory of Materials Physics, Institute of Solid State Physics,
Chinese Academy of Sciences, Hefei 230031, China}
\author{Mei Yan Ni}\affiliation{School
of Electronic Science and Applied Physics, Hefei University of
Technology, Hefei 230009, China}
\author{Guo Ren Zhang}
\author{Wang Huai Zhou}
\author{Yong Gang Li}
\author{Xiao Hong Zheng}
\author{Zhi Zeng}\thanks{Email: zzeng@theory.issp.ac.cn}\affiliation{Key Laboratory of Materials Physics, Institute of
Solid State Physics, Chinese Academy of Sciences, Hefei 230031,
China}

\begin{abstract}
The thermodynamic, kinetic and magnetic properties of the hydrogen
monomer on doped graphene layers were studied by ab initio
simulations. Electron doping was found to heighten the diffusion
potential barrier, while hole doping lowers it. However, both kinds
of dopings heighten the desorption potential barrier. The underlying
mechanism was revealed by investigating the effect of doping on the
bond strength of graphene and on the electron transfer and the
coulomb interaction between the hydrogen monomer and graphene. The
kinetic properties of H and D monomers on doped graphene layers
during both the annealing process (annealing time $t_0 =$300 s) and
the constant-rate heating process (heating rate $\alpha =$1.0 K/s)
were simulated. Both electron and hole dopings were found to
generally increase the desorption temperatures of hydrogen monomers.
Electron doping was found to prevent the diffusion of hydrogen
monomers, while the hole doping enhances their diffusion.
Macroscopic diffusion of hydrogen monomers on graphene can be
achieved when the doping-hole density reaches $5.0\times10^{13}$
cm$^{-2}$. The magnetic moment and exchange splitting were found to
be reduced by both electron and hole dopings, which was explained by
a simple exchange model. The study in this report can further
enhance the understanding of the interaction between hydrogen and
graphene and is expected to be helpful in the design of
hydrogenated-graphene-based devices.
\end{abstract}


\maketitle

\section{INTRODUCTION}
\par In recent years, there have been many investigations into
hydrogenated graphite surfaces and graphene due to their importance
in astronomic
exploration,\cite{coey,hornekaer96,hornekaer97,ferro368} nuclear
industry,\cite{ferro368,morris} graphite (graphene)-based hydrogen
storage\cite{anthony} and graphene-based electronic
devices.\cite{shytov,ryu,elias,sofo75,balog131,bostwick103,guisinger9,luo3,lebegue79,chen96,wright97,flores20}

\par Study of the thermodynamic and kinetic properties of hydrogen
adatoms on graphene can help to understand the interaction mechanism
between the hydrogen adatom and the graphite surface (graphene
layer). The usability of the newly proposed
hydrogenated-graphene-based devices relies heavily on these
properties. Many experiments have been carried out on the
thermodynamic and kinetic properties of hydrogen adatoms on graphite
surfaces,\cite{hornekaer96,hornekaer97,zecho42,zecho117}
substrate-supported graphene layers (monolayer and
multilayer)\cite{guisinger9,elias,ryu,balog131,luo3} and even the
free standing graphene layer.\cite{elias,meyer454} The hydrogenation
of graphene starts with the adsorption of a hydrogen monomer, which
breaks an aromatic $\pi$ bond in graphene and makes the adsorption
of other hydrogen atoms very easy. Thus, the thermodynamic and
kinetic properties of the hydrogen monomer are essential to the
hydrogenation process, which are having been investigated
theoretically in recent
years.\cite{hornekaer97,ferro120,casolo130,herrero79,roman45,boukhvalov77,boukhvalov21,huang}
The ab initio simulation from our previous paper\cite{huang} has
given some predictions of the kinetic properties of the hydrogen
monomer on neutral graphene, which closely reproduced some
experimental observations\cite{hornekaer97,baouche125}. Furthermore,
the hydrogen monomer has been used by some theorists to engineer the
electronic structures, magnetic properties and transport properties
of some graphene-based devices.\cite{bang81,soriano81} The
thermodynamic and kinetic properties of the hydrogen monomer are
critical to the stability of these devices.

\par The charge doping of graphene is a common phenomenon in
intercalated graphite compounds,\cite{csanyi1,mazin95,calandra74}
metallic surface-supported graphene
layers\cite{martoccia101,giovannetti101,khomyakov79} and gated
graphene layers.\cite{novoselov306,das3} Charge (electron or hole)
doping was expected to influence the thermodynamic and kinetic
properties of hydrogen adatoms (including monomers) on graphene. In
addition, the magnetism in graphene-based materials has recently
drawn tremendous interest from the scientific community due to their
lower density compared to transition metals, compatibility with
biological systems, plasticity and so on,\cite{yazyev73} which
introduces another set of applications for charge-doping effects on
the hydrogenated graphene. Overall, charge doping provides a
possible approach to modifying various properties of
hydrogenated-graphene-based materials and devices to fulfill the
requirements for various applications.

\par In this report, the thermodynamic and kinetic properties of
the hydrogen monomer on doped graphene layers were simulated using a
composite method consisting of density functional theory
(DFT),\cite{kohn140} density functional perturbation theory
(DFPT)\cite{baroni73} and harmonic transition state theory
(hTST).\cite{eyring3,hanggi62,pollak15} The mechanism of the
charge-doping effect on the thermodynamic and kinetic properties of
the hydrogen monomer was revealed by investigating the electronic
structures and phonon spectra. The kinetic properties of H and D
monomers on doped graphene layers during both the annealing process
($t_0 =$300 s) and the constant-rate heating (1.0 K/s) were
simulated. The effect of charge doping on the magnetic properties
derived by DFT calculations have been well explained by a simple
exchange model.

\section{METHODOLOGY}
\par The adsorption energy of the hydrogen monomer $E_{ads}$ is defined as the energy difference
between the totally desorbed state and the adsorbed state
\begin{eqnarray}{\label{E_ads}}
E_{ads}=E^H+E^{GL}-E^{GL+H}
\end{eqnarray}
where $E^H$, $E^{GL}$ and $E^{GL+H}$ are the total energies of an
isolated hydrogen atom, an isolated graphene layer and a graphene
layer with a hydrogen monomer adsorbed on it, respectively. Electron
doping will make the anti-bonding $\pi^*$ orbital in graphene
occupied by electrons, while hole doping will reduce the electronic
occupation of the bonding $\pi$ orbital. Thus, both the bonding
$\pi$ and the anti-bonding $\pi^*$ orbitals, which are together
described by a total $\Pi$ bond, are considered here. The
chemisorption of a hydrogen monomer will break an aromatic $\Pi$
bond. The breaking energy of a $\Pi$ bond ($E_\Pi$) is related to
the adsorption energy of the hydrogen monomer in the chemisorption
state, according to Ferro's analysis.\cite{ferro78} Therefore, a
simple model is used to reexpress $E_{ads}$ as
\begin{eqnarray}{\label{E_ads2}}
E_{ads}(\sigma)=E_{C-H}(\sigma)-E_{\Pi}(\sigma)
\end{eqnarray}
where $\sigma$ is the doping-charge density and $E_{C-H}(\sigma)$ is
the formation energy of a C--H bond. If $E_{C-H}$ is assumed to be
unchanged under charge doping, the contribution of $E_\Pi(\sigma)$
to $E_{ads}(\sigma)$ is explicitly addressed in this way. The
adsorption energy based on this assumption is expressed as
\begin{eqnarray}{\label{E_ads_*}}
E_{ads}^*(\sigma)=E_{C-H}(0)-E_{\Pi}(\sigma)
\end{eqnarray}
Comparing $E_{ads}$ in Equ. (\ref{E_ads2}) with $E_{ads}^*$ in Equ.
(\ref{E_ads_*}) gives the relative contributions of
$E_{C-H}(\sigma)$ and $E_{\Pi}(\sigma)$ to $E_{ads}(\sigma)$.
$E_{\Pi}$ can be calculated from the adsorption energies of the
hydrogen monomer and para-dimer\cite{ferro78}
\begin{eqnarray}{\label{Pi_bond}}
E_{\Pi}(\sigma)=E_{ads}^{para}(\sigma)-2E_{ads}(\sigma)
\end{eqnarray}
where $E_{ads}^{para}$ is the adsorption energy of a hydrogen
para-dimer on graphene.

\par The over-barrier jump frequency $v$ between two local
minimum states (initial and final states or reactant and product
states) is calculated from the quantum-mechanically modified
hTST,\cite{huang,eyring3,toyoura78,sundell76} in the Arrhenius form
of
\begin{eqnarray}{\label{arrhenius}}
v=v^*_{qm}\exp{(-\frac{E_{ac}}{k_BT})}
\end{eqnarray}
where $v^*_{qm}$ is the exponential prefactor and $E_{ac}$ is the
activation energy. The activation energy $E_{ac}$ is defined as the
vibrational zero-point energy corrected potential barrier, which is
expressed as
\begin{eqnarray}{\label{activation_E}}
E_{ac} & = & {\Delta}V_{p}+\frac{1}{2}\sum_{i=1}^{3N-1}\hbar{\omega}_i^S-\frac{1}{2}\sum_{i=1}^{3N}\hbar{\omega}_i^I \nonumber\\
       & = & {\Delta}V_{p}-{\Delta}F_{vib}(0)
\end{eqnarray}
where ${\Delta}V_p$ is the potential barrier in the reaction path,
which can be obtained from DFT calculations; ${\omega}^I_i$ and
${\omega}^S_i$ are the vibrational frequencies of the $i$th mode in
the initial and saddle-point (SP) states, respectively, which can be
obtained from DFPT calculations; ${\Delta}F_{vib}(0)$ is the
vibrational zero-point energy correction. The total vibrational
degrees of freedom are 3N. An imaginary vibrational mode along the
migration coordinate in the SP state is excluded from calculation.
Thus, there are 3N-1 vibrational modes considered for the SP state.
The quantum-mechanically modified prefactor is expressed
as\cite{huang}
\begin{eqnarray}{\label{prefactor}}
v^*_{qm}=\frac{k_BT}{h}\frac{\prod\limits_{i=1}^{3N-1}\exp{(\frac{\hbar\omega_i^S}{k_BT})}\bar
n_T(\omega_i^S)}{\prod\limits_{i=1}^{3N}\exp{(\frac{\hbar\omega_i^I}{k_BT})}\bar
n_T(\omega_i^I)}
\end{eqnarray}
where $\bar n_T(\omega_i)$ is the bosonic phonon occupation number
of the $i$th vibrational mode.

\par The first-order rate equation for the desorption of the hydrogen monomer
is defined as\cite{huang,baouche125,hanggi62}
\begin{eqnarray}
\frac{dn(t)}{dt}=-{v_{des}(T)}n(t)
\end{eqnarray}
where $v_{des}$ is the desorption jump frequency of the hydrogen
monomer; $n(t)$ is the residual number of hydrogen monomers on a
graphene layer at time $t$; $t=0$ represents the starting time of
the kinetic movement.

\par In the annealing process ($T=T_0$), the variation of the
residual number with respect to the annealing time interval ($t_0$)
is expressed as
\begin{eqnarray}
n(t)=n(0)\exp{[-v_{des}(T_0)t_0]}
\end{eqnarray}
where $n(0)$ is the monomer number at the starting time ($t=0$).
Conceptually, the relative residual monomer number ($n(t)/n(0)$) is
also the desorption probability of a single monomer. In this case,
the diffusional property can be described by the mean square
displacement ($<|{\bf r}(t)-{\bf r}(0)|^2>$) of a monomer parallel
to the graphene layer. From Fick's second law, we have
\begin{eqnarray}
<|{\bf{r}}(t)-{\bf{r}}(0)|^2>=2dD_{ad}(T_0)t_0
\end{eqnarray}
where $d$ is the dimensionality of the diffusion of a hydrogen
monomer on graphene, taken to be 2 here, and $D_{ad}$ is the
temperature-dependent diffusion coefficient of the adatom, described
by\cite{toyoura78}
\begin{eqnarray}
D_{ad}=\frac{1}{2d}{\Gamma}a^2
\end{eqnarray}
where $\Gamma$ is the total jump frequency of the monomer, and $a$
is the jump length. For the diffusion of a hydrogen monomer on
graphene, $\Gamma$ is taken as $3v_{diff}$ ($v_{diff}$ is the
diffusion jump frequency) because there are three equivalent paths
for the diffusion of a monomer on graphene, and $a$ is taken as the
optimized C--C bond length of 1.426 \AA{}. The diffusion radius of
the monomer is defined as the square root of the mean square
displacement
\begin{eqnarray}{\label{r_dif_annealing}}
r_{dif}=\sqrt{2dD_{ad}(T_0)t}
\end{eqnarray}
which directly determines the diffusional mobility of the hydrogen
monomer on graphene.

\par In the constant-rate heating process ($T={\alpha}t$), the variation of the
residual number with respect to time is
\begin{eqnarray}
n(t)=n(0)\exp{[-\int_0^tv_{des}(T)dt]}
\end{eqnarray}
where the heating rate $\alpha$ is always taken as 1.0 K/s in
experiments.\cite{hornekaer96,zecho42,zecho117} In this case, the
diffusion radius is expressed as
\begin{eqnarray}{\label{r_dif_heating}}
r_{dif}=\sqrt{2d\int_0^t{D_{ad}(T)dt}}
\end{eqnarray}

\par
\begin{figure}[p]
\scalebox{0.25}[0.25]{\includegraphics{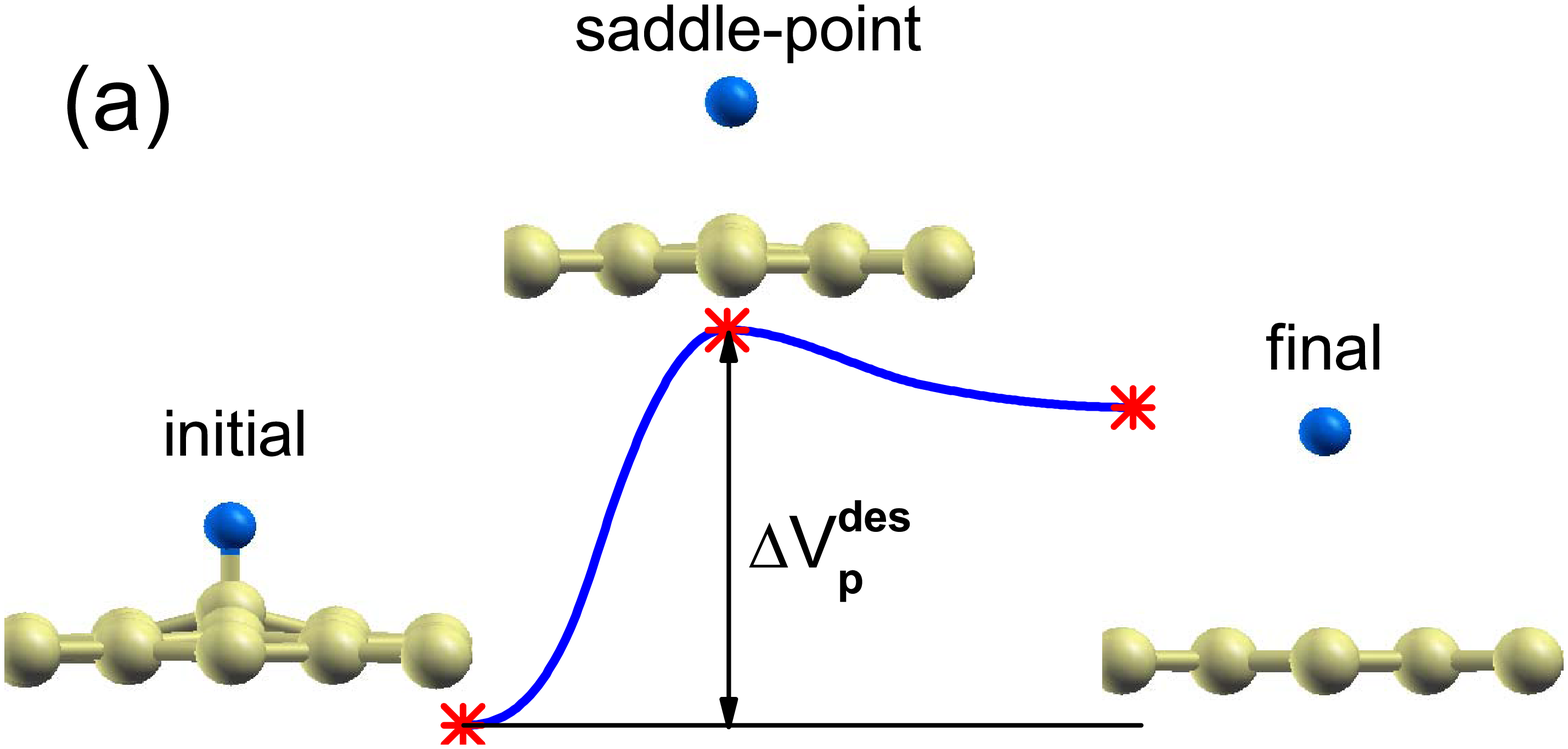}}\\
\scalebox{0.25}[0.25]{\includegraphics{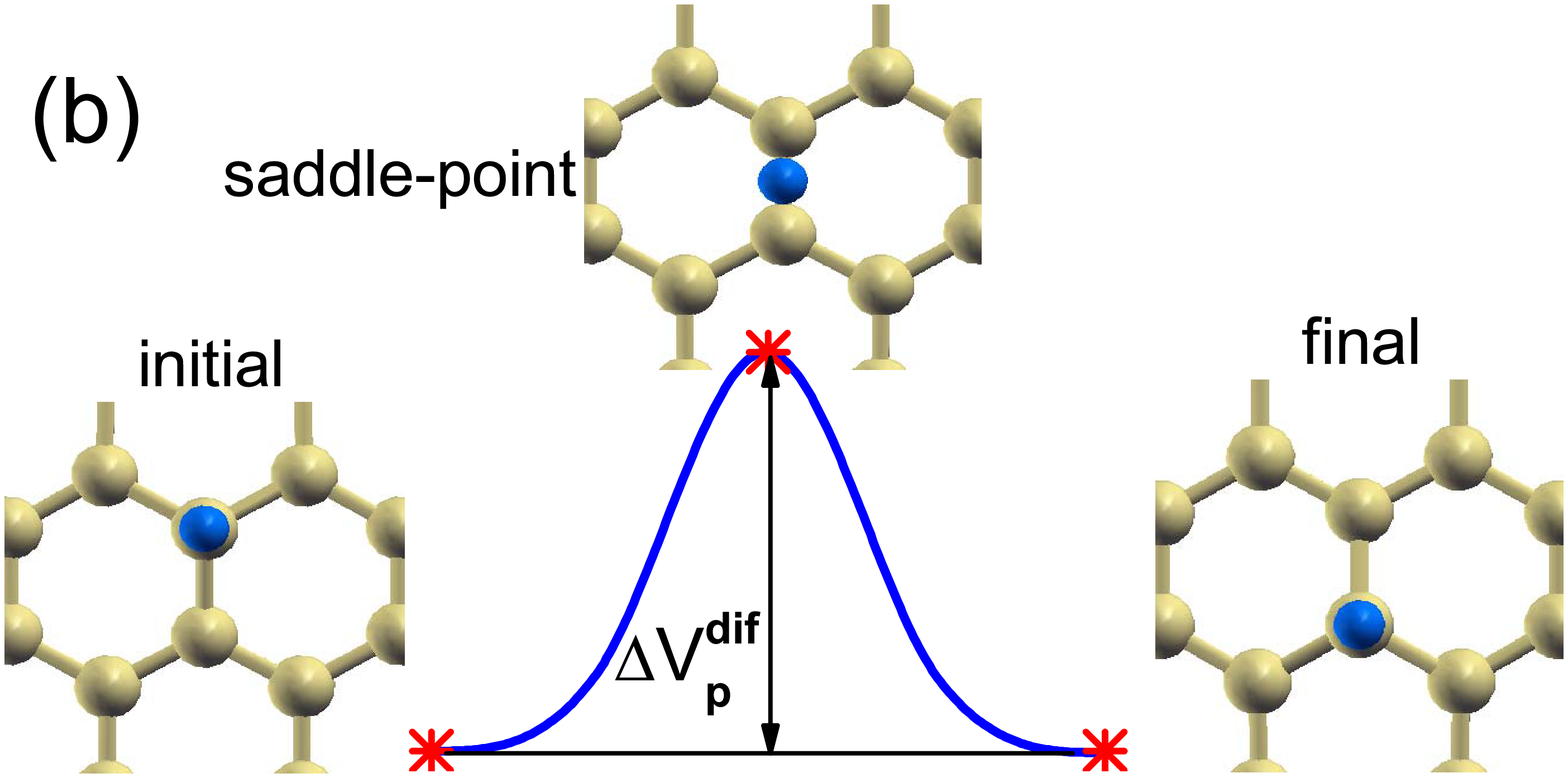}}
\caption{\label{MEPs}(Color online) The MEPs for the (a) desorption
and (b) diffusion of the hydrogen monomer on graphene. The initial,
saddle-point (SP) and final states in the paths are labeled with
stars, and their structures are shown alongside the graph. The
yellow spheres are carbon atoms and the smaller blue spheres are
hydrogen atoms. The final state in the desorption MEP is set to be
the physisorption state.}
\end{figure}

 In this study, a hydrogen monomer on a $5\times5$ periodic supercell of graphene (50 C atoms),
with a 10 \AA{} vacuum along the direction perpendicular to the
surface, was taken as an isolated monomer. The DFT and DFPT
calculations were carried out using the Quantum Espresso code
package.\cite{giannozzi21} The ultrasoft\cite{vanderbilt41}
spin-polarized PBE\cite{perdew77} pseudopotentials were applied to
describe the electronic exchange and correlation energy. The
reaction path was described by the minimum energy path (MEP) between
two local minimum states (Fig. \ref{MEPs}), and the MEPs for the
desorption and diffusion of a hydrogen monomer on graphene were
calculated using the climbing-image nudged elastic band
method.\cite{henkelman113} For calculating the vibrational
frequencies of the periodic supercells, only the gamma point at the
Brillouin zone center is selected. The charge-doped systems were
compensated with the same numbers of opposite background charges.
The optimized C--C bond length of graphene is 1.426 \AA{}; thus, a
charge density of $1.0\times10^{13}$ cm$^{-2}$ corresponds to a
charge number of 0.1314 $e$ added into the graphene supercell. More
details about the computational method and tests can be found in
Ref. [\onlinecite{huang}].

\section{RESULTS AND DISCUSSION}
\par
\begin{figure}[p]
\scalebox{0.20}[0.20]{\includegraphics{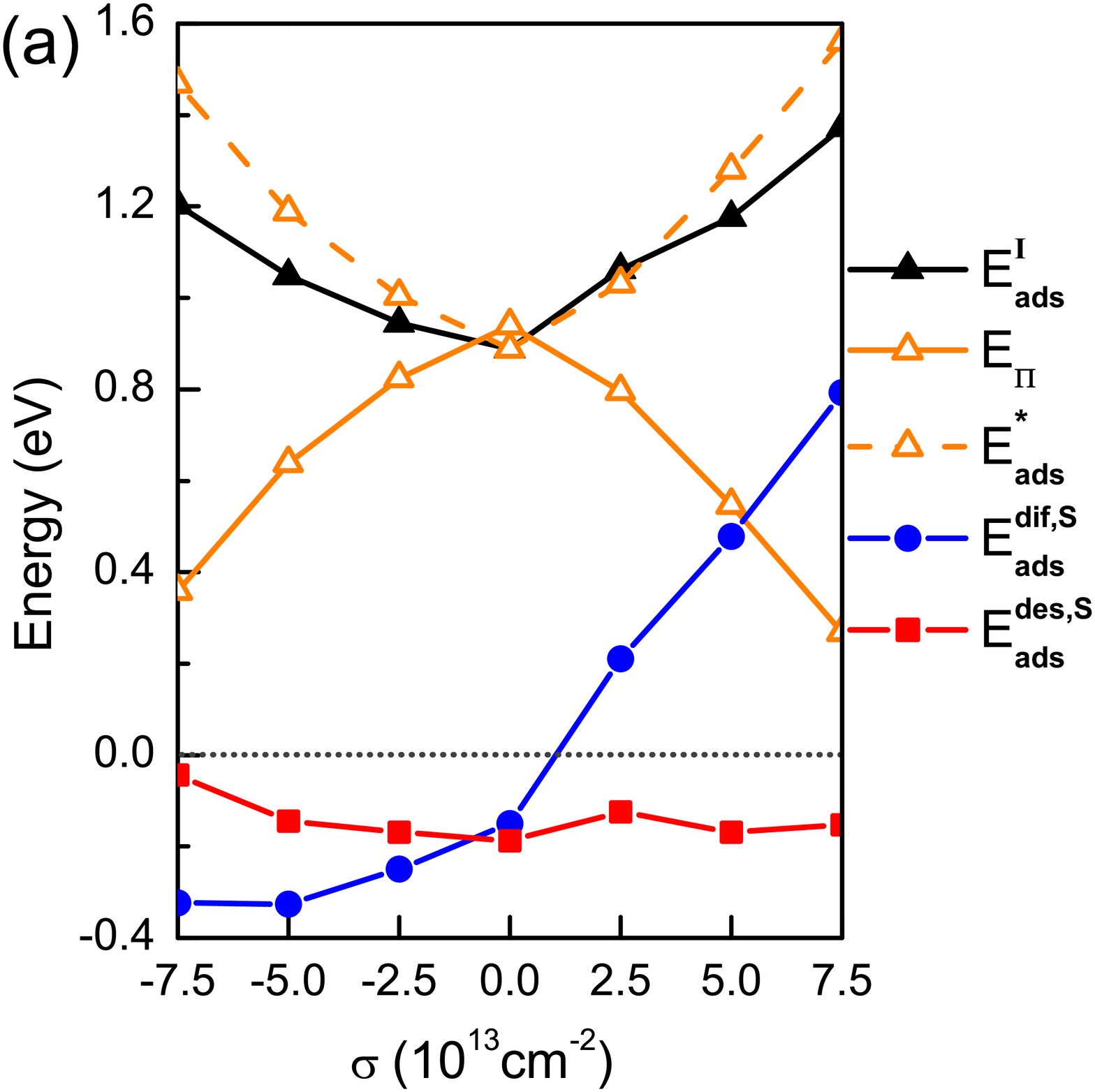}}
\scalebox{0.20}[0.20]{\includegraphics{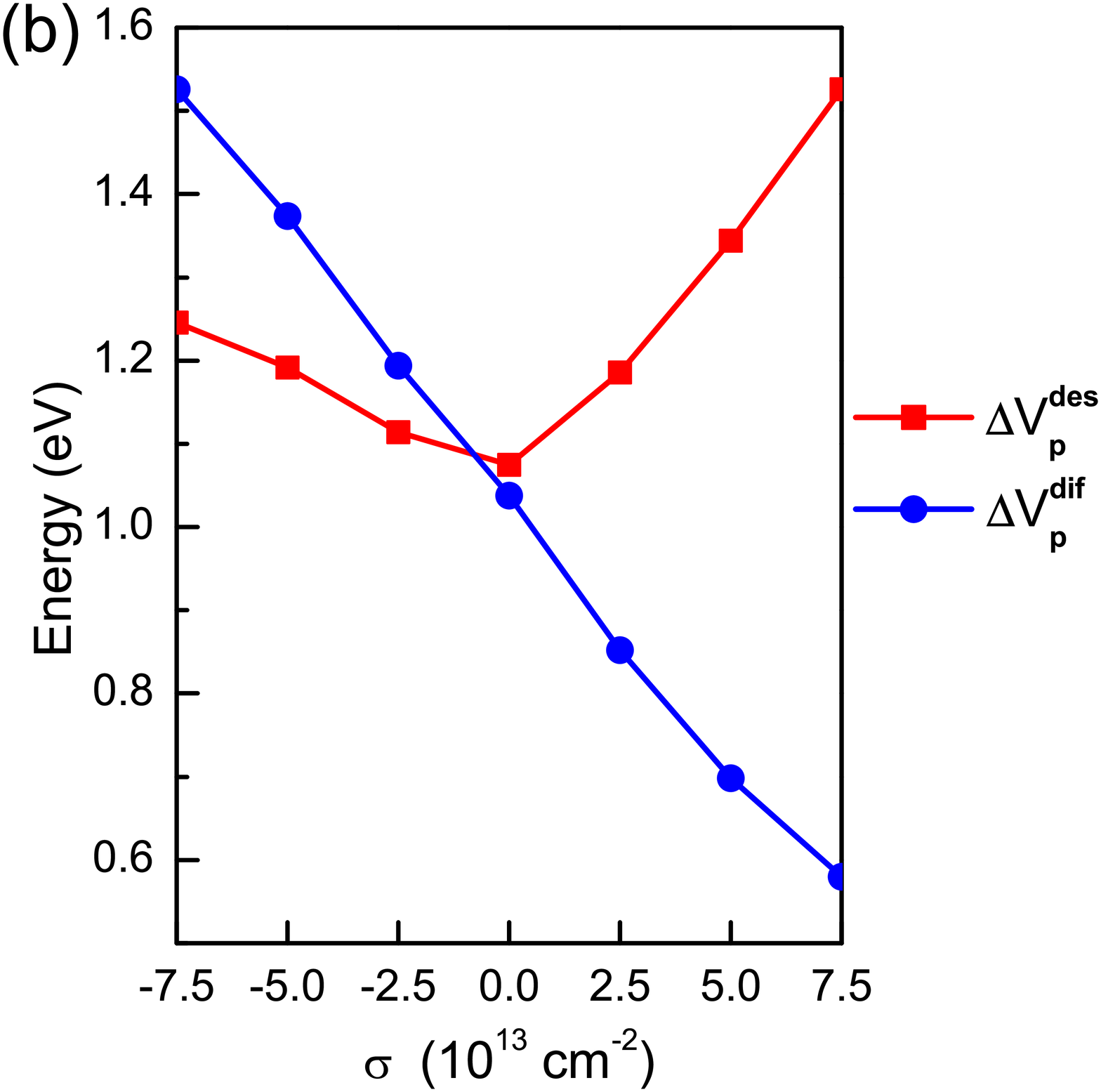}}
\caption{\label{ads_kinetic_E}(Color online) (a) The variations of
$E_{ads}^I$, $E_{ads}^{des,S}$, $E_{ads}^{dif,S}$, $E_\Pi$ and
$E_{ads}^*$ with respect to $\sigma$. (b) The variations of $\Delta
V_p^{des}$ and $\Delta V_p^{dif}$ with respect to $\sigma$.}
\end{figure}

 Schematic drawings of the MEPs for the desorption and diffusion processes of a hydrogen
monomer on graphene are shown in Fig. \ref{MEPs}, where the
structures of the initial (reactant), SP (transition) and final
(product) states are also shown. The initial state is the
chemisorption state, and the final state here is set as the
physisorption state. The potential barrier is the energy difference
between the SP state and the initial state. The variation of the
adsorption energy of the hydrogen monomer in the initial state
($E_{ads}^I$), diffusion SP state ($E_{ads}^{dif,S}$) and desorption
SP state ($E_{ads}^{des,S}$), and of the potential barriers for the
desorption ($\Delta V_p^{des}$) and diffusion ($\Delta V_p^{dif}$)
of the hydrogen monomer with respect to the doping-charge density
($\sigma$) are shown in Fig. \ref{ads_kinetic_E}.

\par $E_{ads}^I$ increases with the number of
doping electrons (negative charge) or doping holes (positive
charge). However, the curve of $E_{ads}^I$ is asymmetric about
$\sigma = 0.0$, because the rate of increase is greater under hole
doping than under electron doping. $E_{ads}^{dif,S}$ increases
monotonically with increasing $\sigma$ from negative to positive,
and the rate of increase becomes larger than that of $E_{ads}^I$
under hole doping, but smaller under electron doping. The curve of
$E_{ads}^{des,S}$ is somewhat flat compared with those of
$E_{ads}^{I}$ and $E_{ads}^{dif,S}$. The potential barrier of
diffusion $\Delta{V}_p^{dif}$ ($=E_{ads}^I - E_{ads}^{dif,S}$)
decreases monotonically with increasing $\sigma$ from negative to
positive, while the potential barrier of desorption
$\Delta{V}_p^{des}$ ($=E_{ads}^I - E_{ads}^{des,S}$) increases with
the number of doping electrons or holes. These phenomena in energy
are related to the electronic structure of graphene and to the
electron transfer and coulomb interaction between the hydrogen
monomer and graphene, which will be explained in detail below.

\par
\begin{figure}[p]
\scalebox{0.25}[0.25]{\includegraphics{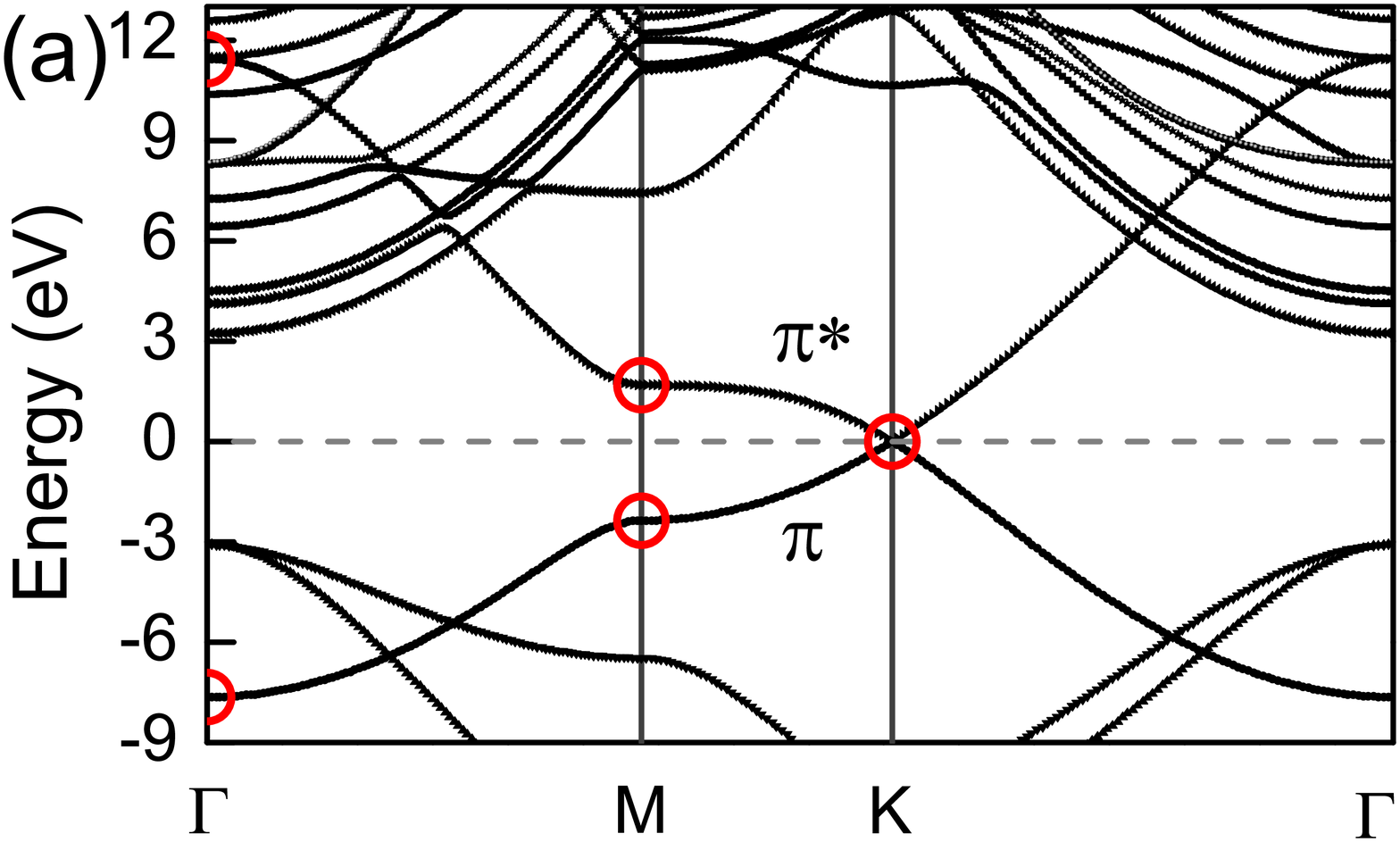}}
\scalebox{0.4}[0.4]{\includegraphics{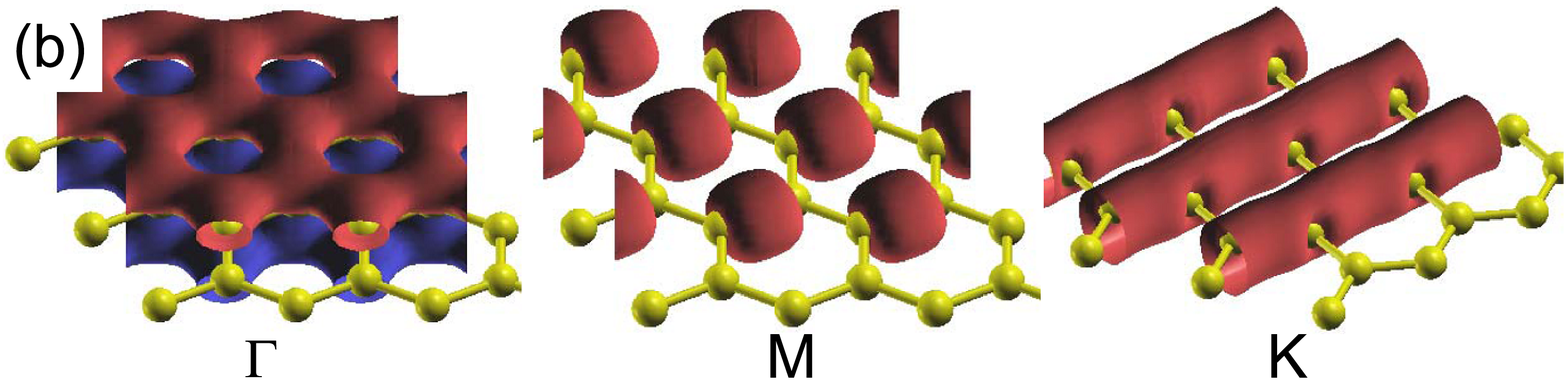}}
\scalebox{0.4}[0.4]{\includegraphics{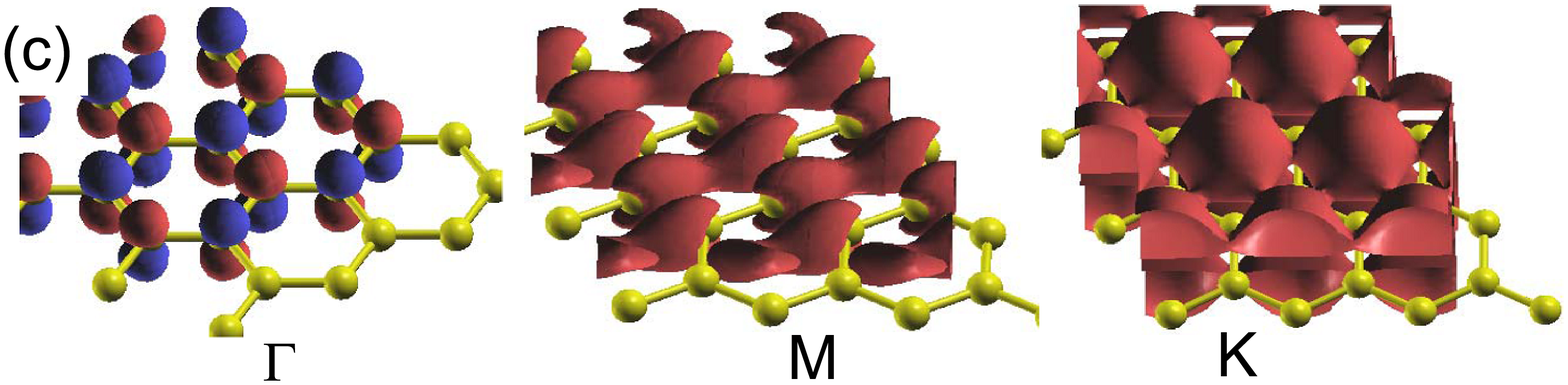}}
\caption{\label{electronic_bands}(Color online) (a) The band
structure of neutral graphene. The orbital shapes in real space of
(b) the bonding $\pi$ band and (c) the anti-bonding $\pi^*$ band at
$\Gamma$, M and K points. In (a), the bonding $\pi$ band and the
anti-bonding $\pi^*$ band at $\Gamma$, M and K points are each
labeled with a red circle.}
\end{figure}

 The band structure of neutral graphene is shown in Fig.
\ref{electronic_bands}(a), where the anti-bonding $\pi^*$ band is
above the Fermi level and the bonding $\pi$ band is below the Fermi
level. The anti-bonding and bonding characteristics of the $\pi^*$
and $\pi$ bands are shown by their orbital shapes in real space in
Fig. \ref{electronic_bands}(b) and (c), respectively. Upon electron
doping, the Fermi level goes up, leading to the occupation of the
anti-bonding $\pi^*$ band, and the strength of the total $\Pi$ bond
is weakened. Upon hole doping, the Fermi level goes down, reducing
the electron-occupation number in the bonding $\pi$ band, and thus
the strength of the $\Pi$ bond is also weakened. The $E_\Pi$
obtained at different $\sigma$ by Equ. \ref{Pi_bond} are shown in
Fig. \ref{ads_kinetic_E}(a). $E_\Pi$ decreases with increasing the
number of doping electrons or holes, inverse to the variation of
$E_{ads}^I$. This is consistent with the weakening of the $\Pi$ bond
predicted from the electronic structure of graphene. The curve of
$E_\Pi$ is asymmetric because the rate of decrease is greater under
hole doping than under electron doping, like the variation of
$E_{ads}^I$. From Equ. \ref{E_ads2} and \ref{E_ads_*}, the weakening
of the $\Pi$ bond, namely the decrease of $E_\Pi$, results in the
increase of $E_{ads}^I$ and $E_{ads}^*$, both of which exhibit a
similar trend (Fig. \ref{ads_kinetic_E}(a)). However, the general
lower value of $E^I_{ads}$ than that of $E^*_{ads}$ indicates the
effect of the charge doping on $E_{C-H}$, which is related to the
electron transfer and coulomb interaction between the hydrogen
monomer and the doped graphene (shown below).

\par
\begin{figure}[p]
\scalebox{0.20}[0.20]{\includegraphics{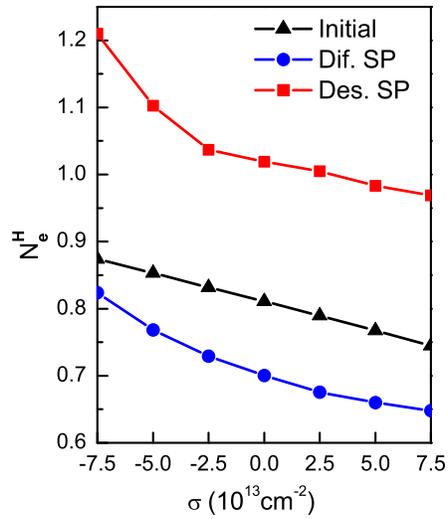}}
\caption{\label{lowdin_charges}(Color online) The variations of
$N_e^H$s in the initial, desorption SP and diffusion SP states with
respect to $\sigma$.}
\end{figure}

\begin{figure}[p]
\scalebox{0.3}[0.3]{\includegraphics{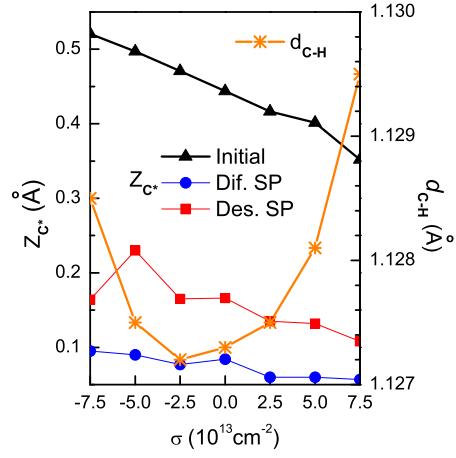}}
\caption{\label{z_c_d_CH}(Color online) The variations of $d_{C-H}$
in the initial state and Z$_{C^*}$s in the initial, desorption SP
and diffusion SP states with respect to $\sigma$.}
\end{figure}

 The calculated L{\"o}wdin electronic populations, also
called the {\itshape{net atomic population}} by
Mulliken,\cite{mulliken23} of the hydrogen monomer ($N^H_e$) in the
initial, diffusion SP and desorption SP states are shown in Fig.
\ref{lowdin_charges}. $N_e^H$ decreases monotonically with
increasing $\sigma$ from negative to positive. The $N_e^H$s for the
initial state are smaller than 1.0 due to the electron transfer from
the hydrogen monomer to graphene, mainly through the orbital
hybridization between them. This orbital hybridization in the
initial state will be shown later to be unchanged by the charge
doping. However, the electron transfer is still influenced by charge
doping due to the chemical-potential difference between the
hybridized hydrogen monomer and graphene. The monotonic decrease of
$N_e^H$ for the initial state indicates that the electron doping
tends to block but the hole doping to promote the electron transfer
from the adsorbed hydrogen monomer to graphene. This blockage of
electron transfer under electron doping tends to reduce the affinity
of graphene to bond with the hydrogen monomer, so that $E_{C-H}$
decreases with increasing doping electrons, making $E_{ads}^I <
E_{ads}^*$ under electron doping (by Equ. (\ref{E_ads2}) and
(\ref{E_ads_*})). On the other hand, hole doping tends to promote
electron transfer so as to increase $E_{C-H}$, making $E_{ads}^I$ a
little larger than $E_{ads}^*$ at $\sigma \lesssim 2.5\times10^{13}$
cm$^{-2}$. However, at $\sigma > 2.5\times10^{13}$ cm$^{-2}$, the
further promoted electron transfer increases the coulomb repulsion
between the positively charged hydrogen monomer and the hole-doped
graphene, which weakens the strength of the C--H bond and makes
$E_{ads}^I < E_{ads}^*$ again. The weakening of the strength of the
C--H bond by charge doping can also be qualitatively observed from
the variation of its bond length ($d_{C-H}$) under charge doping in
Fig. \ref{z_c_d_CH}. The variation of the protrusion height of the
H-bonded C atom ($Z_{C^*}$) with $\sigma$ is also shown in Fig.
\ref{z_c_d_CH}, which will be used later. It should be noted that
the increase of $E_{ads}^I$ by charge doping is mainly determined by
the redution of the breaking energy of a total $\Pi$ bond ($E_\Pi$),
as described in detail in the previous paragraph, and the electron
transfer and coulomb repulsion play minor roles. However, from the
electronic density of states (DOS) of the neutral desorption and
diffusion SP states as described in Ref. \onlinecite{huang}, the
hydrogen monomer hybridizes little with graphene. This is the same
for these doped cases, but only with Fermi levels tuned by charge
doping (not shown). Thus, $E_{ads}^{des,S}$ and $E_{ads}^{dif,S}$
should be not significantly related to $E_{\Pi}$. The $N_e^H$s for
the desorption SP state are larger than 1.0 under electron doping
and smaller than 1.0 under hole doping. The doping charges
distribute across the system according to the requirement of
electrostatic equilibrium, which explains the flatness of the
$E_{ads}^{des,S}(\sigma)$ curve in Fig. \ref{ads_kinetic_E}(a). The
$N_e^H$s for the diffusion SP state are significantly smaller than
1.0. From the DOS in Ref. \onlinecite{huang}, the hydrogen monomer
will dope graphene with some number of itinerant electrons in this
state. The monotonic decrease of $N_e^H$ (promotion of the electron
transfer) with increasing $\sigma$ is due to the chemical-potential
difference between the hydrogen monomer and doped graphene, which is
the same as the electronic population of the lithium adatom on
charged carbon nanotubes.\cite{myni35} This promotion of electron
transfer results in the increase of the affinity of graphene to bond
with the hydrogen monomer in the diffusion SP state and results in
the increase of $E_{ads}^{dif,S}$. As a result, the increase of
$\Delta{V}_p^{des}$ ($= E_{ads}^I - E_{ads}^{des,S}$) under charge
(electron or hole) doping and the monotonic decrease of
$\Delta{V}_p^{dif}$ ($= E_{ads}^I - E_{ads}^{dif,S}$) with
increasing $\sigma$ are conjunctly due to the effects of charge
doping on the strength of the total $\Pi$ bond of graphene, the
electron transfer and to the coulomb interaction between the
hydrogen monomer and graphene.

\par
\begin{figure}[p]
\scalebox{0.25}[0.25]{\includegraphics{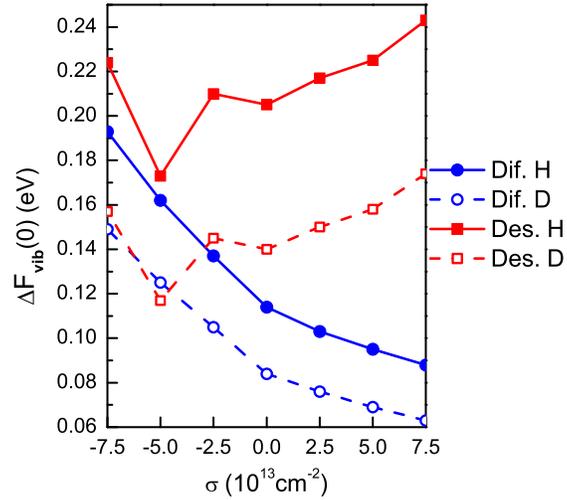}}
\caption{\label{ZPEs}(Color online) The variations of the
$\Delta{F}_{vib}(0)$s for the desorption and diffusion of H and D
monomers on graphene with respect to $\sigma$.}
\end{figure}

 The calculated vibrational zero-point
energy corrections ($\Delta{F}_{vib}(0)$) for the desorption and
diffusion of H and D monomers under various $\sigma$s are shown in
Fig. \ref{ZPEs}. The isotope effect of $\Delta{F}_{vib}(0)$ is
obvious; it decreases with increasing monomer mass. This will result
in the reversed isotope effect in the activation energy by Equ.
\ref{activation_E}. $\Delta{F}_{vib}(0)$ and its isotopic difference
for desorption are generally larger than those for diffusion. These
vibrational properties are related to the spectra of the localized
vibrational modes of H and D monomers, which has been discussed in
detail in Ref. \onlinecite{huang}. The $\Delta{F}_{vib}(0)$s for the
diffusion of H and D monomers decrease with increasing $\sigma$, the
same as $\Delta{V}_p^{dif}$. This is because the decrease of the
potential barrier $\Delta{V}_p^{dif}$ will result in the decrease of
the diffusion-MEP potential curvature around the initial state,
which then results in the lowering of the frequencies of the
effective vibrations for the diffusion of H and D monomers whose
zero-point energies equal their $\Delta{F}_{vib}(0)$s. Except for
the kinks at $\sigma$ of $-5.0\times10^{13}$ cm$^{-2}$, the
$\Delta{F}_{vib}(0)$s for the desorption of H and D monomers present
the same variation as ${\Delta}V_p^{des}$ with respect to $\sigma$,
which is that the values increase with the number of doping
electrons or holes. This is related to the increase of the potential
barrier of the desorption-MEP and the increase of the frequencies of
effective vibrations for the desorption of H and D monomers. The
kinks in the $\Delta{F}_{vib}(0)$ curves for the desorption of H and
D monomers at $\sigma$ of $-5.0\times10^{13}$ cm$^{-2}$ are related
to the structure of the desorption SP state. The protrusion height
of the H-bonded $C^*$ ($Z_{C^*}$) in the chemisorption and SP states
can be used to determine the structural information. It can be seen
in Fig. \ref{z_c_d_CH}, there is an obvious kink in the
$Z_{C^*}(\sigma)$ curve for the desorption SP state at $\sigma =
-5.0\times10^{-13}$ cm$^{-2}$, while the $Z_{C^*}(\sigma)$ curves
for the initial and diffusion SP states are much smoother. This kink
makes the $C^*$ closer to the hydrogen monomer in the desorption SP
state at $\sigma = -5.0\times10^{-13}$ cm$^{-2}$ than at other
charge densities. Then, the interaction between hydrogen monomer and
graphene in the desorption SP state at $\sigma = -5.0\times10^{-13}$
cm$^{-2}$ will be larger than at other charge densities, which also
can be reflected in the spectra of the localized vibrational modes
of the hydrogen monomer (shown below).

\par
\begin{figure}[p]
\scalebox{0.4}[0.4]{\includegraphics{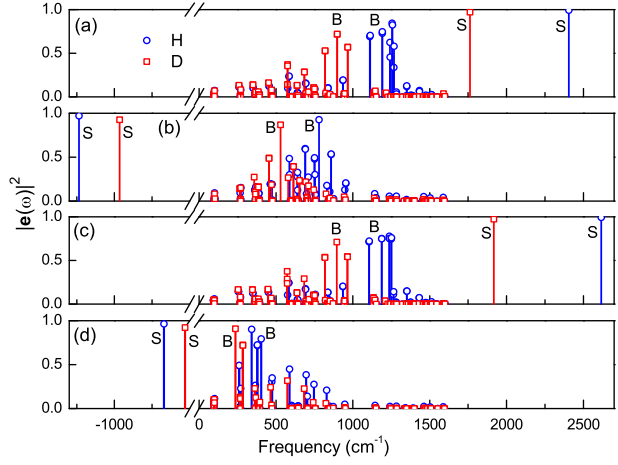}}
\caption{\label{phonons}(Color online) The calculated spectra of
$|{\bf{e}}(\omega_i)|^2$ ({\itshape{i}} is the index of the
vibrational mode) of H (D) monomer in (a) the initial state and (b)
the desorption SP state at $\sigma$ of $-5.0\times10^{13}$
cm$^{-2}$, and in (c) the initial state and (d) the desorption SP
state at $\sigma$ of $-7.5\times10^{13}$ cm$^{-2}$.}
\end{figure}

 The localized vibrational modes have large displacements
(${\bf{e}}(\omega_i)$, {\itshape{i}} is the index of the vibrational
mode) of the hydrogen monomer in their eigenvectors of the
vibrational dynamic matrix. The spectra of $|{\bf{e}}(\omega_i)|^2$
for the initial and desorption SP states at $\sigma$s of
$-5.0\times10^{13}$ and $-7.5\times10^{13}$ cm$^{-2}$ are shown in
Fig. \ref{phonons}. The stretching (S) modes become imaginary in the
desorption SP states. Compared with the bending (B) modes in the
initial states, the bending modes in the desorption SP states are
much less lowered at $\sigma$ of $-5.0\times10^{13}$ cm$^{-2}$ than
at $\sigma$ of $-7.5\times10^{13}$ cm$^{-2}$. The phonon spectra at
other $\sigma$s (e.g. see the Fig. 3 in Ref. \onlinecite{huang} for
the neutral case) are close to those at $\sigma$ of
$-7.5\times10^{13}$ cm$^{-2}$ and shift smoothly with $\sigma$,
which is also reflected by the $\Delta{F}_{vib}(0)$ curves (Fig.
\ref{ZPEs}). As described in the previous paragraph, the interaction
between the hydrogen monomer and graphene in the desorption SP state
at $\sigma$ of $-5.0\times10^{13}$ cm$^{-2}$ is larger than those at
other $\sigma$s, and thus (the absolute values of) the vibrational
frequencies of the localized vibrational modes of the hydrogen
monomer are higher at this $\sigma$ than those at other $\sigma$s.
By Equ. \ref{activation_E}, the result is that the
$\Delta{F}_{vib}(0)$s for the desorption SP state are lower at
$\sigma$ of $-5.0\times10^{13}$ cm$^{-2}$ than at other $\sigma$s,
and that the $E_{ac}$s for desorption at $\sigma$s of $-5.0$ and
$-7.5\times10^{13}$ cm$^{-2}$ are very close to each other. By Equ.
\ref{prefactor} and from the detailed analysis in Ref.
\onlinecite{huang}, the stiffening of the localized vibrational
modes in the desorption SP state at $\sigma$ of $-5.0\times10^{13}$
cm$^{-2}$ will also result in the decrease of the corresponding
$v_{qm}^*$s compared with those at other $\sigma$s.

\par
\begin{figure}[p]
\scalebox{0.15}[0.15]{\includegraphics{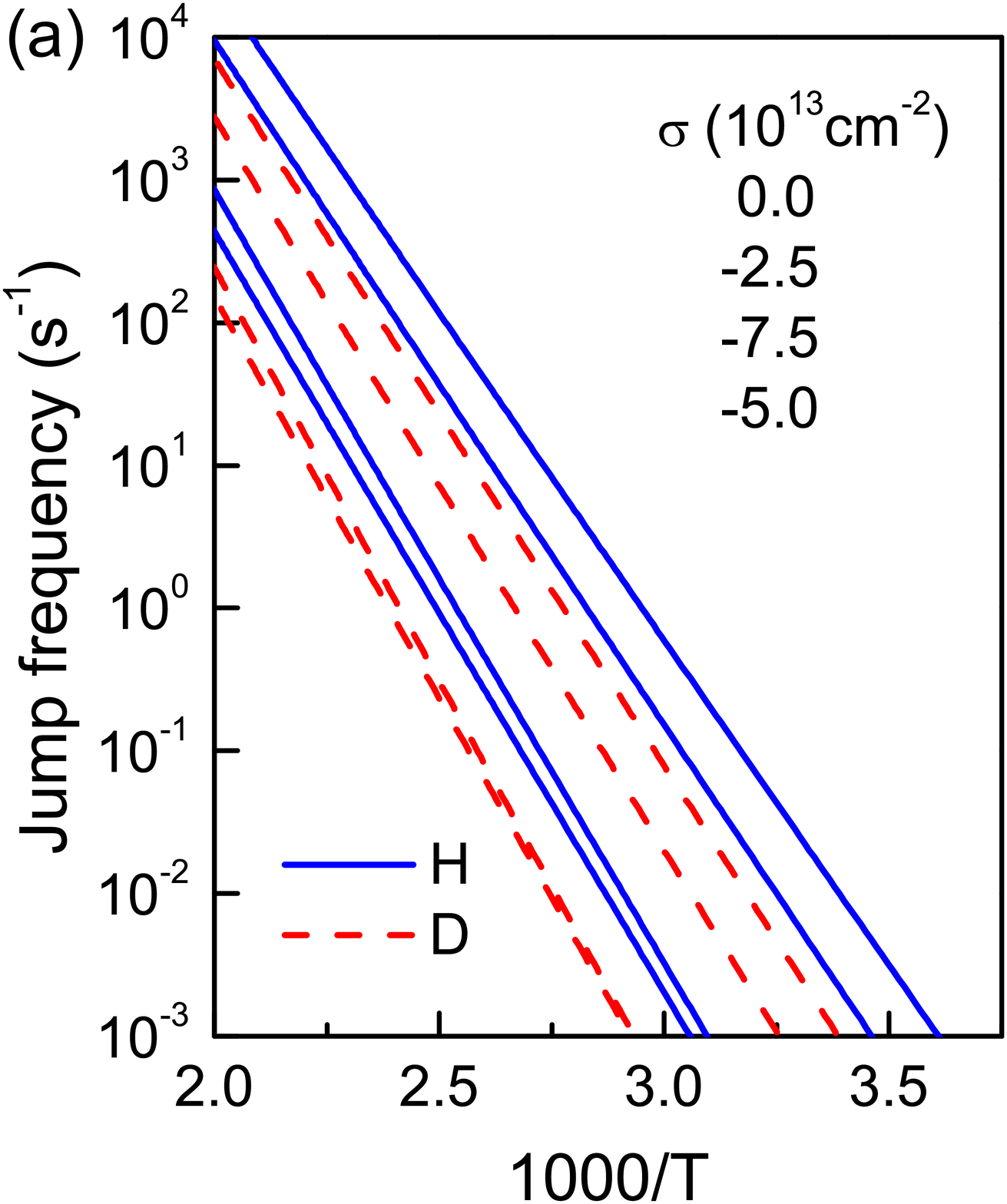}}
\scalebox{0.15}[0.15]{\includegraphics{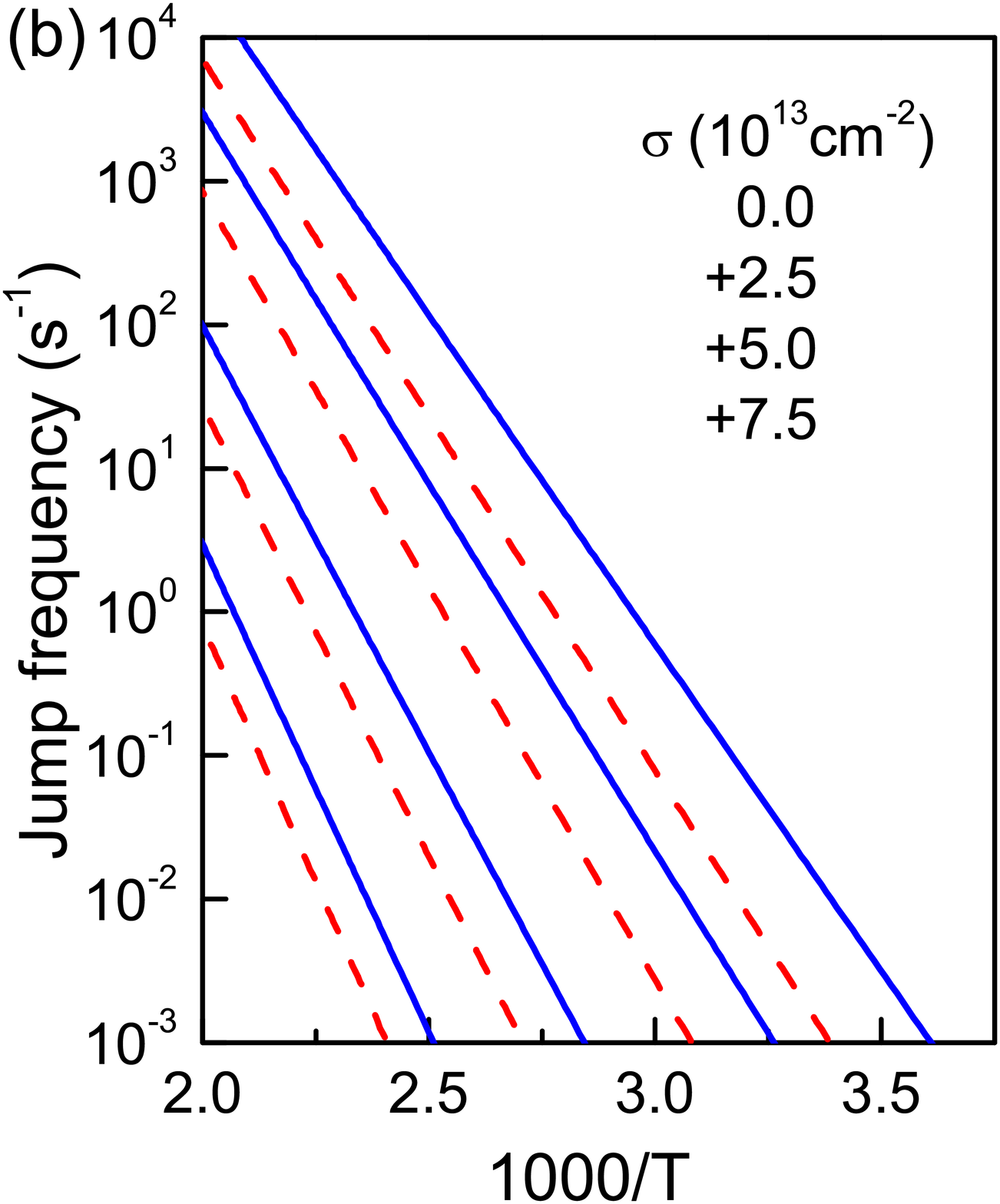}}
\scalebox{0.15}[0.15]{\includegraphics{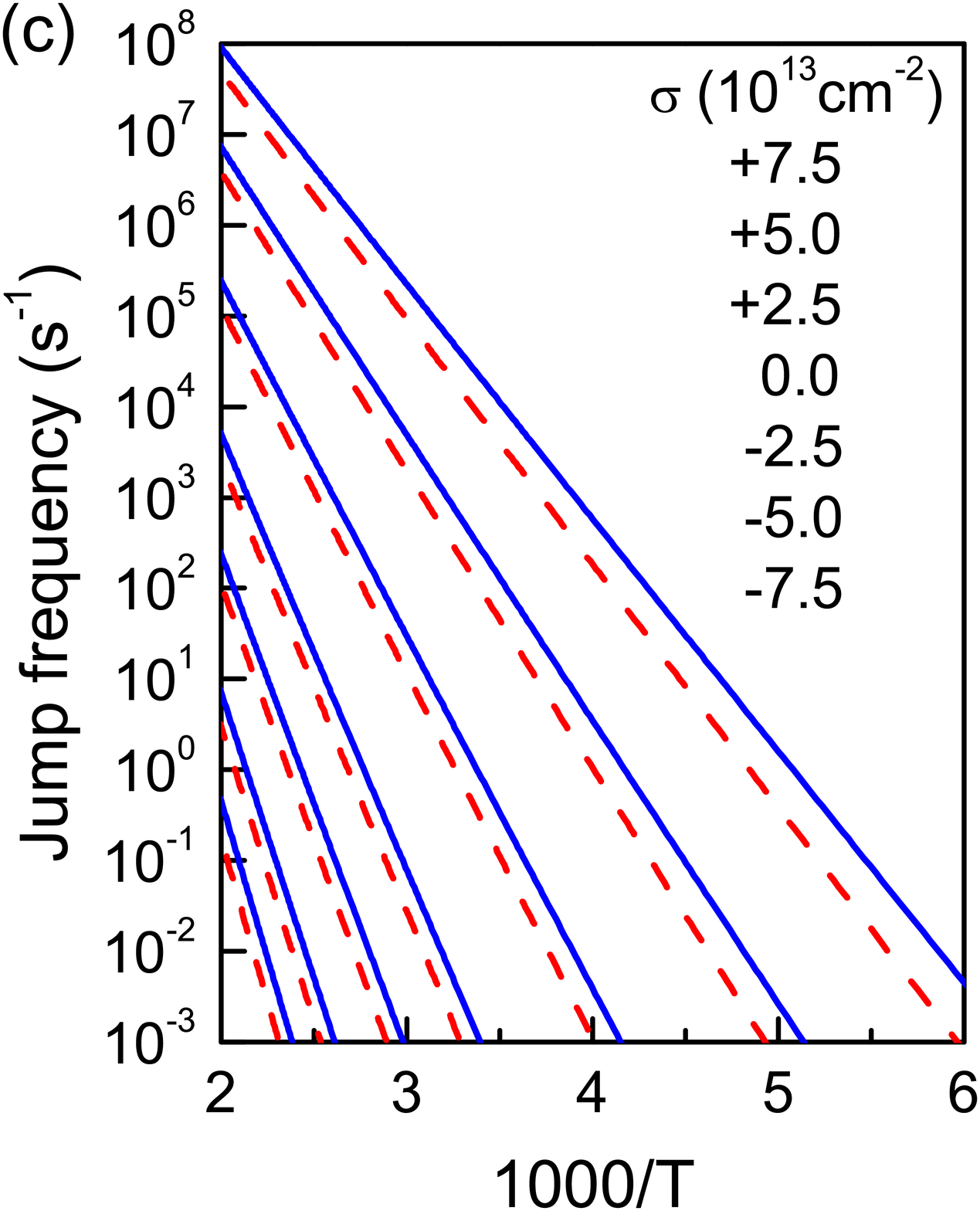}}
\caption{\label{jumpfreq}(Color online) The variations of the jump
frequencies for the (a, b) desorption and (c) diffusion of H and D
monomers on graphene at various $\sigma$s with respect to the
inverse of temperature.}
\end{figure}

 The calculated jump frequencies ($v$) for the desorption and diffusion of H and
D monomers on doped graphene are shown in Fig. \ref{jumpfreq}. The
value of $v$ decreases with increasing monomer mass. This isotope
effect is due to the isotope effect in $\Delta{F}_{vib}(0)$ (Fig.
\ref{ZPEs}) by Equ. \ref{arrhenius} and \ref{activation_E}, which
has been discussed in detail in Ref. \onlinecite{huang}. In Fig.
\ref{jumpfreq}(a) and (b), the desorption $v$ generally decreases
with increasing the number of doping electrons (negative $\sigma$)
or holes (positive $\sigma$), with the exception that the $v$ for
the desorption of H (D) monomer at $\sigma$ of $-5.0\times10^{13}$
cm$^{-2}$ is less than that at $\sigma$ of $-7.5\times10^{13}$
cm$^{-2}$. By Equ. \ref{arrhenius}, $v$ is exponentially determined
by $E_{ac}(\sigma)$ ($= \Delta{V}_p^{des} + \Delta{F}_{vib}(0)$),
where the linear dependence on $v_{qm}^*$ usually plays a minor
role. Except at $\sigma$ of $-5.0\times10^{13}$ cm$^{-2}$, the
$E_{ac}$ for the desorption of H (D) monomer increases with the
number of doping electrons or holes; thus, the desorption $v$
generally decreases with increasing the number of doping electrons
or holes. However, the lowering of the $\Delta{F}_{vib}(0)$s at
$\sigma$ of $-5.0\times10^{13}$ cm$^{-2}$ makes the $E_{ac}$s very
close to those at $\sigma$ of $-7.5\times10^{13}$ cm$^{-2}$, and the
relative magnitudes of the $v$s at these two charge densities are
determined by the values of their $v_{qm}^*$s. The $v_{qm}^*$ at
$\sigma$ of $-5.0\times10^{13}$ cm$^{-2}$ is smaller than those at
other $\sigma$s due to the stiffening of the localized modes (not
shown), as described in the previous paragraph. In Fig.
\ref{jumpfreq}(c), the diffusion $v$ increases monotonically with
$\sigma$, which is due to the monotonic decrease of $E_{ac}$ for
diffusion with increasing $\sigma$. Thus, it can be concluded that
any kind of charge (electron or hole) doping will make the bonding
between the hydrogen monomer and graphene kinetically more stable,
and electron doping will prevent but the hole doping will trigger
the diffusion of the hydrogen monomer on graphene. Thus, when
increasing $\sigma$ from negative to positive, there should be a
crossover between the priorities of desorption and diffusion of H
(D) monomer on graphene, like the crossover between
$\Delta{V}_p^{des}$ and $\Delta{V}_p^{dif}$ in Fig.
\ref{ads_kinetic_E}.

\par
\begin{figure}[p]
\scalebox{0.22}[0.22]{\includegraphics{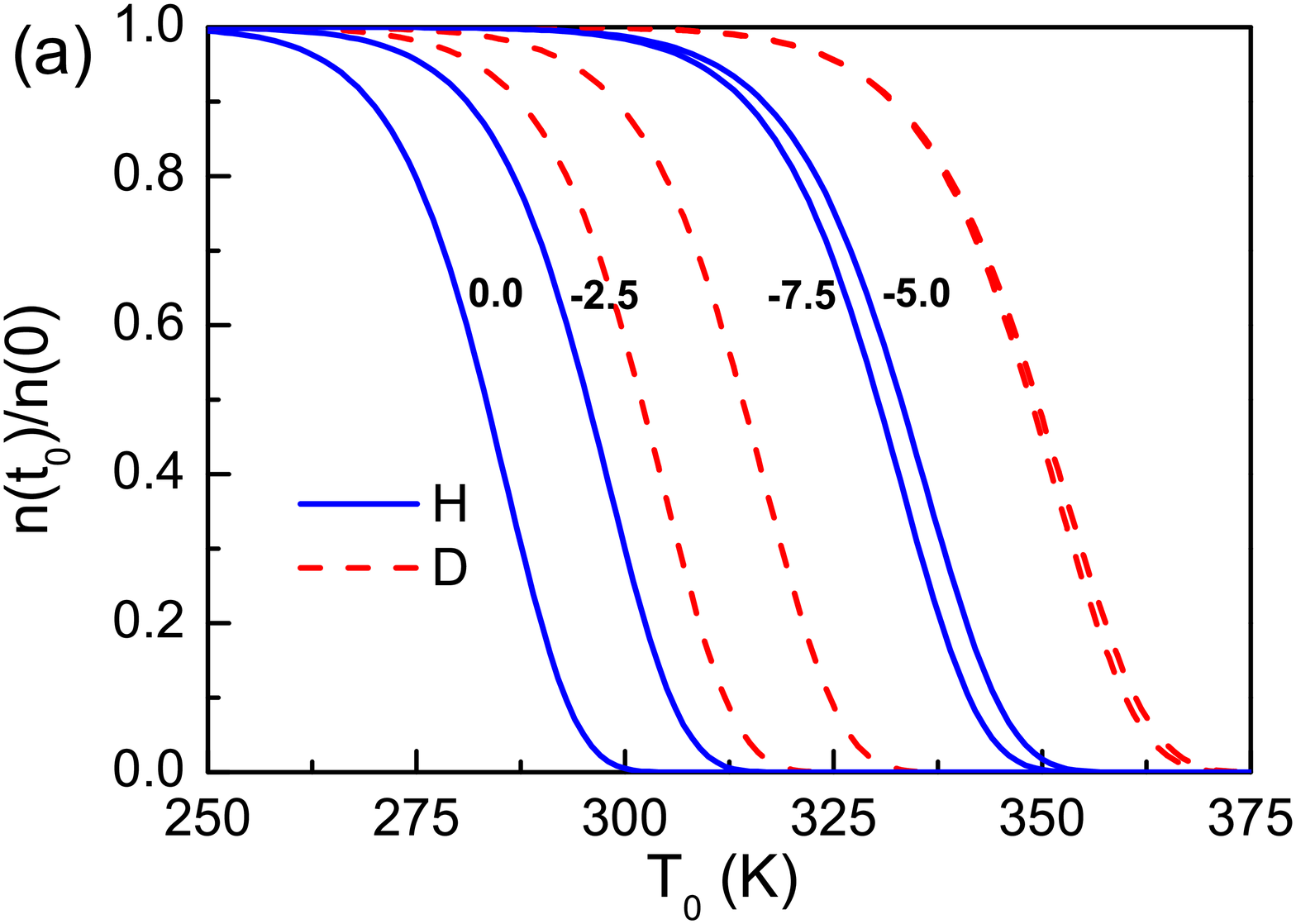}}
\scalebox{0.25}[0.25]{\includegraphics{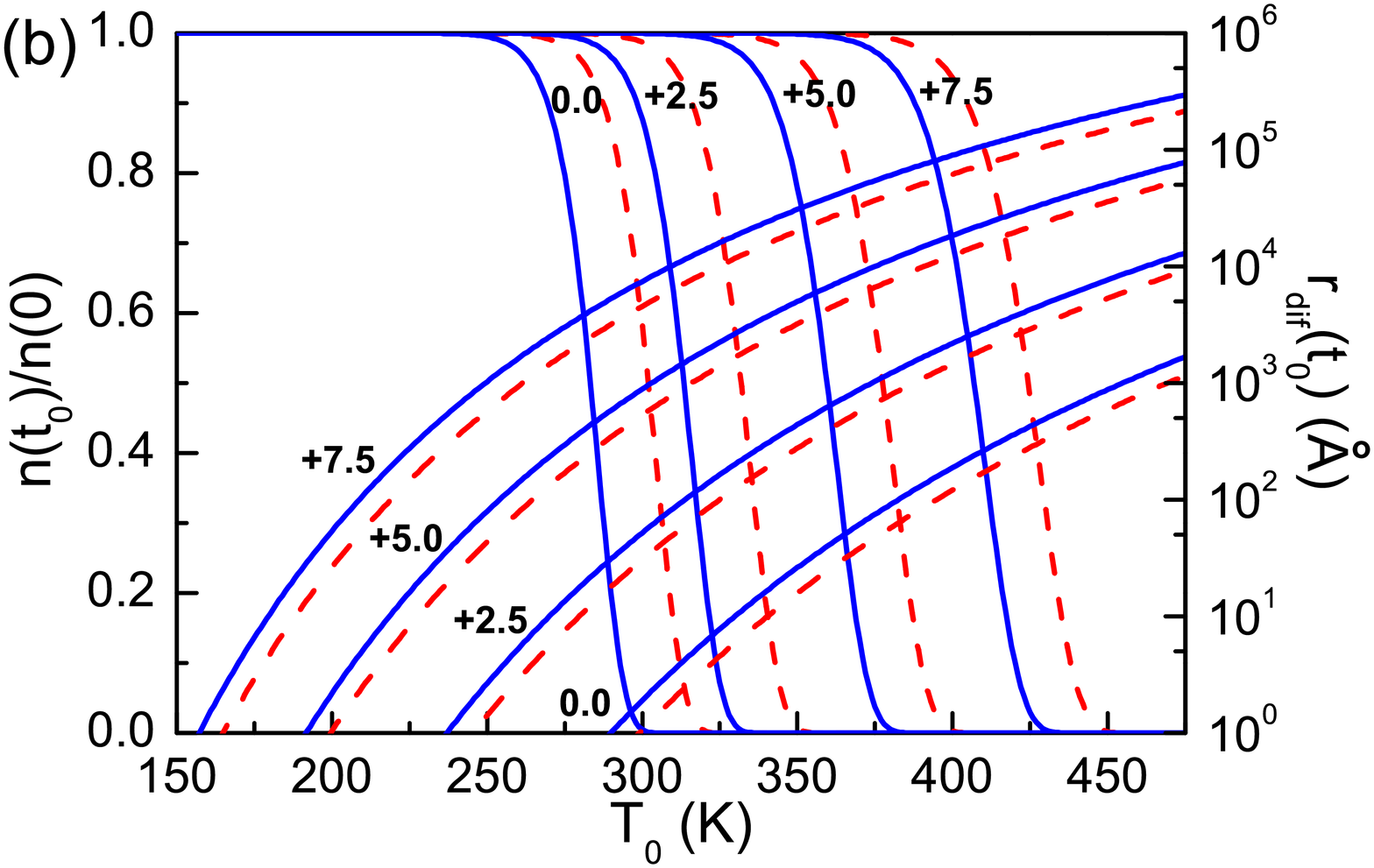}}
\caption{\label{anneal}(Color online) The variations of the
$n(t_0)/n(0)$s and $r_{dif}(t_0)$s for H and D monomers at various
$\sigma$s with respect to the annealing temperature ($T_0$) in the
annealing process ($t_0=300 s$). The $r_{dif}(t_0)$s at negative
$\sigma$s are not shown, because the diffusion of H (D) monomer is
stopped by electron doping.}
\end{figure}

 In the annealing process, the system is kept at an annealing
temperature ($T_0$) for a fixed time interval ($t_0$). In the
simulation here of the kinetic properties of H and D monomers in the
annealing process, $t_0$ was set to be 300s. For electron-doped
graphene layers, only the desorption of the hydrogen monomer was
considered in the simulation, because the diffusion is stopped by
electron doping. However, for hole-doped graphene layers, both the
desorption and the diffusion of the hydrogen monomer were
considered. The properties of desorption and diffusion can be
characterized by the relative residual monomer number (or the
desorption probability) ($n(t_0)/n(0)$) and the diffusion radius
($r_{dif}(t_0)$), respectively. The calculated variations of the
$n(t_0)/n(0)$s and $r_{dif}(t_0)$s of H and D monomers at various
$\sigma$s with respect to $T_0$ are shown in Fig. \ref{anneal}. At
$n(t_0)/n(0)=0.5$, the corresponding $T_0$ was defined to be the
desorption temperature. The desorption temperature of H (D) on
neutral graphene is 282 (301) K. Under electron doping, the
desorption temperatures of H (D) monomer are 295 (314), 333 (349)
and 330 (348) K at $\sigma$s of $-2.5$, $-5.0$ and
$-7.5\times10^{13}$ cm$^{-2}$, respectively. Under hole doping, the
desorption temperatures are 313 (331), 359 (376) and 407 (424) K at
$\sigma$s of 2.5, 5.0 and $7.5\times10^{13}$ cm$^{-2}$,
respectively. The desorption temperatures for D monomer are about 18
K higher than those for H monomer. The $r_{dif}$ of H (D) monomer on
neutral graphene at the desorption temperature is 0.6 (1.0) \AA{},
which is smaller than the C--C bond length of 1.426 \AA{} and
indicates the immobility of the hydrogen monomer in diffusion. Under
hole doping, the $r_{dif}$s of H (D) monomer at the desorption
temperatures are 9.0 (12.5) $n$m, 0.65 (0.71) $\mu$m and 10.7 (9.8)
$\mu$m at $\sigma$s of 2.5, 5.0 and $7.5\times10^{13}$ cm$^{-2}$,
respectively. The isotope effect on the desorption and diffusion of
the hydrogen monomer can be concluded to be that the lighter
hydrogen monomer is desorbed and diffuses more easily. The
variations of the desorption temperatures and $r_{dif}$s with
respect to $\sigma$ are the same as those of the $v$s for desorption
and diffusion, respectively. If diffusion with radius above 0.1
$\mu$m is defined to be macroscopic diffusion, the macroscopic
diffusion of H (D) monomer at temperatures below the desorption
temperature can be achieved when $\sigma$ reaches $5.0\times10^{13}$
cm$^{-2}$.

\par
\begin{figure}[p]
\scalebox{0.22}[0.22]{\includegraphics{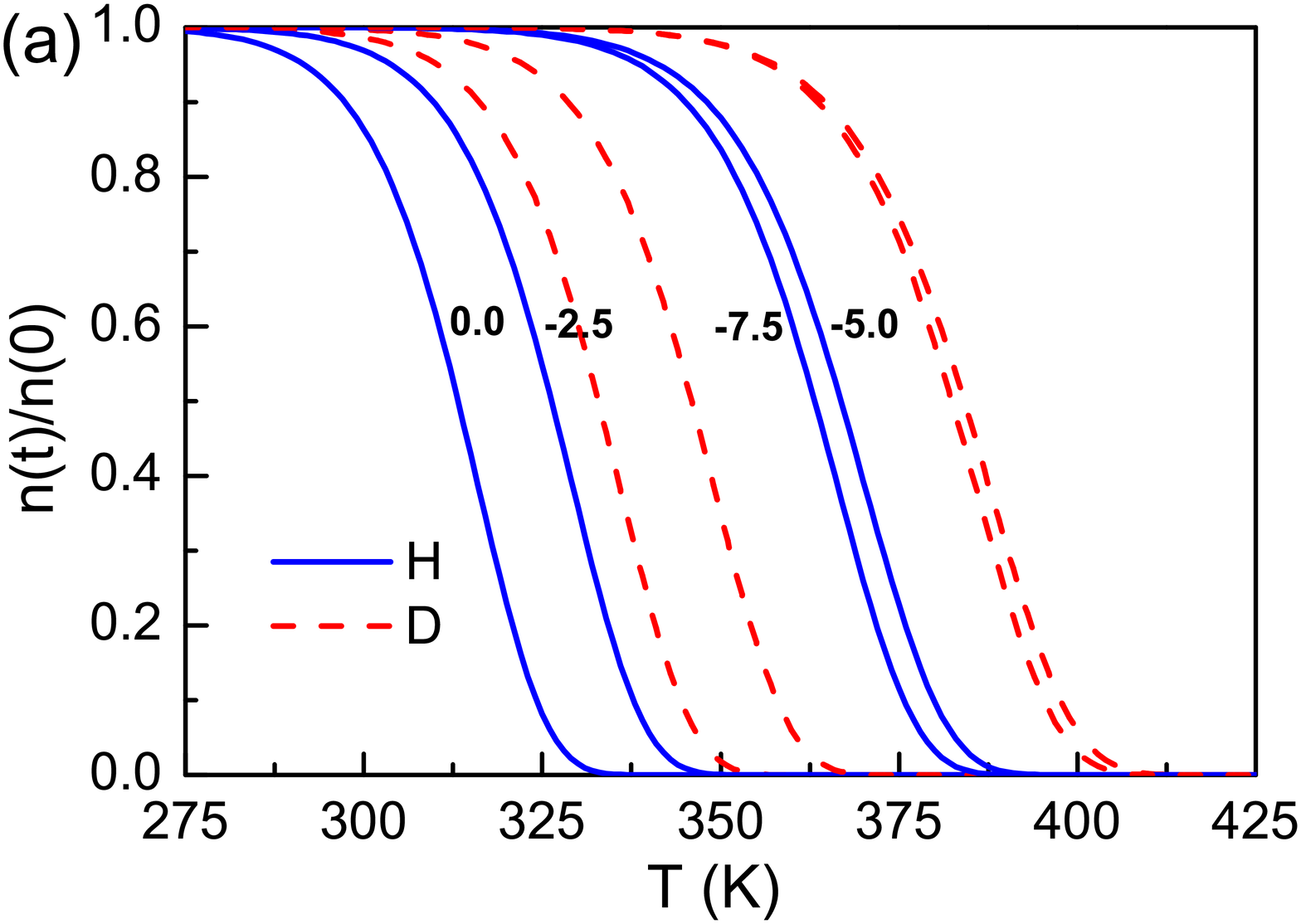}}
\scalebox{0.25}[0.25]{\includegraphics{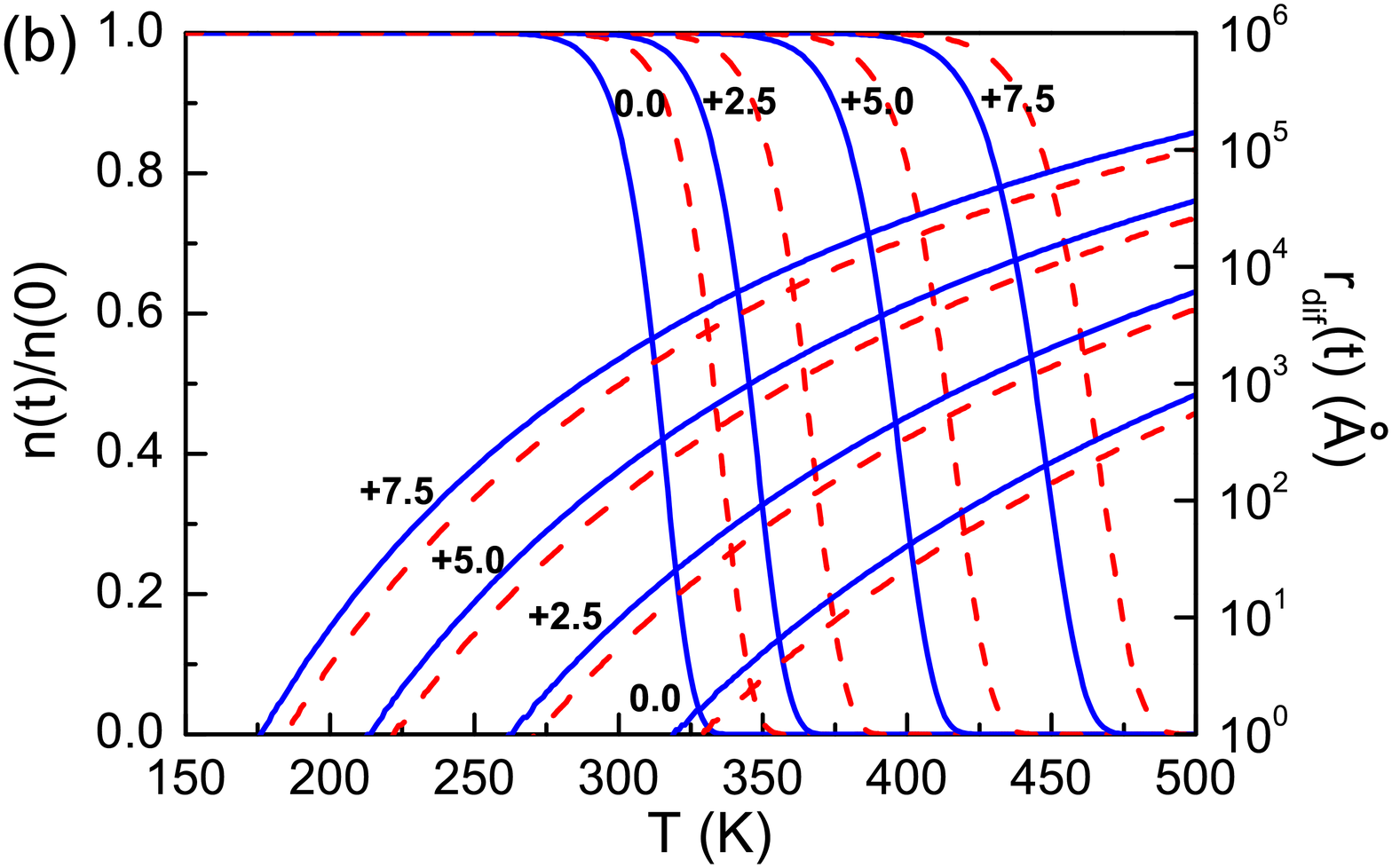}}
\caption{\label{TDS}(Color online) The variations of the
$n(t)/n(0)$s and $r_{dif}(t)$s for H and D monomers at various
$\sigma$s with respect to temperature in the constant-rate heating
process ($\alpha=$ 1.0 K/s). The $r_{dif}(t)$s at negative $\sigma$s
are not shown, because the diffusion of H (D) monomer is stopped by
electron doping.}
\end{figure}

 In the constant-rate heating process, the system is heated at a constant rate
($T=\alpha t$). The heating rate ($\alpha$) was taken to be 1.0 K/s
in the simulation here, which was a commonly used value in
experiments. The calculated variations of the $n(t_0)/n(0)$s and
$r_{dif}(t_0)$s at various $\sigma$s are shown in Fig. \ref{TDS}.
The desorption temperature of H (D) on neutral graphene is 313 (333)
K. Under electron doping, the desorption temperatures of H (D)
monomer are 326 (346), 367 (384) and 363 (382) K at $\sigma$s of
$-2.5$, $-5.0$ and $-7.5\times10^{13}$ cm$^{-2}$, respectively.
Under hole doping, the desorption temperatures are 345 (363), 394
(412) and 445 (461) K at $\sigma$s of 2.5, 5.0 and
$7.5\times10^{13}$ cm$^{-2}$, respectively. The desorption
temperatures in the constant-rate heating process are about 33 K
higher than those in the annealing process. The desorption
temperatures for D monomer are also about 18 K higher than those for
H monomer. The $r_{dif}$ of H (D) monomer on neutral graphene at the
desorption temperature is 0.8 (1.3) \AA{}, which is also smaller
than the C--C bond length and indicates the immobility of the
hydrogen monomer in diffusion. Under hole doping, the $r_{dif}$s of
H (D) monomer at the desorption temperatures are 7.5 (9.6) $n$m,
0.41 (0.41) $\mu$m and 5.9 (5.6) $\mu$m at $\sigma$s of 2.5, 5.0 and
$7.5\times10^{13}$ cm$^{-2}$, respectively. The isotope effect on
the desorption and diffusion of the hydrogen monomer on graphene and
the variations of the desorption temperatures and the $r_{dif}$s
with respect to $\sigma$ are the same as those in the annealing
process. The macroscopic diffusion of the hydrogen monomer is also
achieved in this constant-rate heating process when $\sigma$ reaches
$5.0\times10^{13}$ cm$^{-2}$.

\par Although the diffusion radius can be increased by hole doping,
the hydrogen monomer on heavily hole-doped graphene layers may not
be easily observed in experiments, because highly diffusive hydrogen
monomers tend to meet and form hydrogen dimers and clusters, which
are much more stable than isolated monomers,\cite{hornekaer97} or
quickly diffuse to the edge of the finite-sized sample. Thus, to
observe the diffusion of a hydrogen monomer experimentally, a medium
hole density was suggested. The hydrogen monomer on suspended
graphene sheet has been detected to be stable by transmission
electron microscopy (TEM) at room temperature.\cite{meyer454}
Although this stability may be due to the low actual sample
temperature caused by the cold trap used in the experiment, the
effect of the incident electrons in TEM should also be considered.
The incident electrons will inevitably excite the electrons in the
bonding $\pi$ band up to the anti-bonding $\pi^*$ band in graphene.
This excitation reduces the electronic occupation number of the
$\pi$ band and increases that of the $\pi^*$ band together, which
weakens the total $\Pi$ bond in graphene and enhances the stability
of the hydrogen monomer on graphene, according to the analysis
above.

\par
\begin{figure}[p]
\scalebox{0.3}[0.3]{\includegraphics{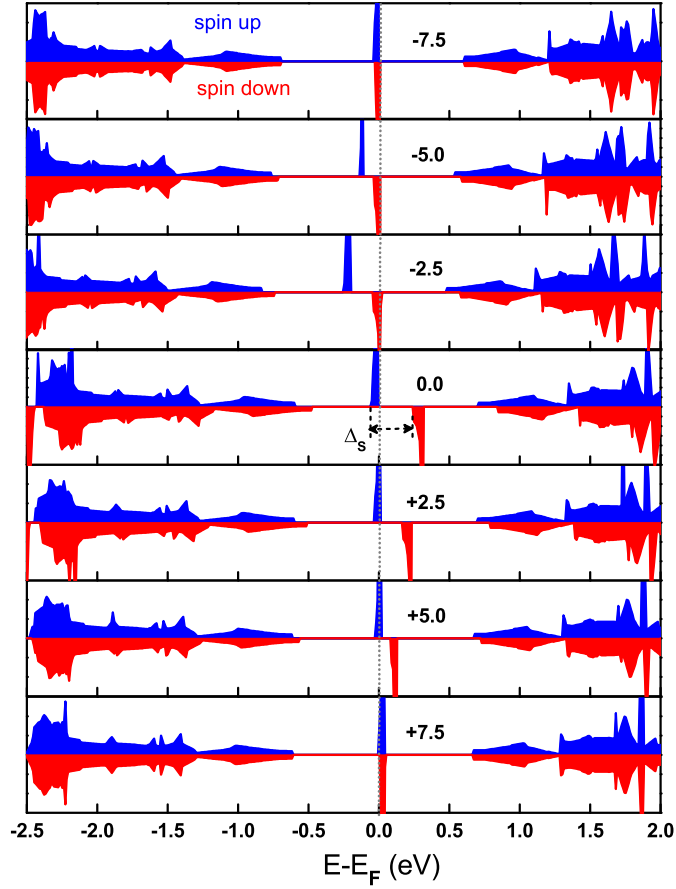}}
\caption{\label{DOS_all}(Color online) The electronic DOS for the
chemisorption states at various $\sigma$s. The gray vertical dotted
line at the Fermi level guides the eyes. The definition of
$\Delta_s$ is shown in the DOS spectrum at $\sigma = 0.0$.}
\end{figure}

\begin{figure}[p]
\scalebox{0.3}[0.3]{\includegraphics{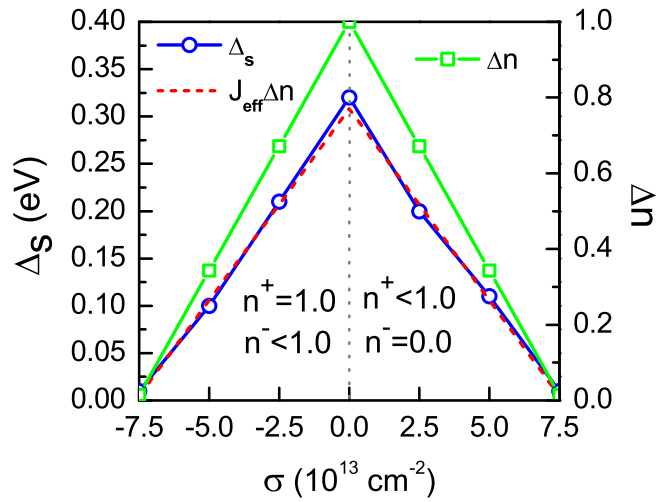}}
\caption{\label{DOS_model}(Color online) The variation of $\Delta$n,
$\Delta_s$ and $\frac{1}{4}$J$_{eff}\Delta$n with respect to
$\sigma$. J$_{eff}$ is fitted to be 1.23 eV.}
\end{figure}

 The adsorption of hydrogen monomer onto neutral
graphene can break an aromatic $\pi$ bond, which results in two
dangling C($p_z$) orbitals. Then, one C($p_z$) orbital bonds with
the  H($1s$) orbital, and the C($p_z$) orbital left unsaturated
forms an occupied spin-polarized quasilocal state (spin-up) around
the hydrogen monomer.\cite{casolo130,huang,verges81} In the spectrum
of the electronic DOS(Fig. \ref{DOS_all}), the occupied quasilocal
state presents as a narrow spin-up peak within the gap between the
conduction and valence bands, and there is another narrow peak above
the Fermi energy (E$_F$) that corresponds to the unoccupied
spin-down quasilocal state. The energy difference between these two
peaks is due to the exchange interaction of electrons, and this
exchange splitting in energy was defined as $\Delta_s$. The
electronic occupation number of these two quasilocal states can be
changed by charge doping. The calculated DOS of the chemisorption
state at various $\sigma$s are shown in Fig. \ref{DOS_all}. The
contribution of the H($1s$) orbital to the total DOS (not shown)
does not significantly vary with charge doping, and is nearly the
same as that in the neutral case in Ref. \onlinecite{huang}, which
indicates that charge doping does not significantly influence the
orbital hybridization between the hydrogen monomer and graphene.
This can also be validated from the invariance of the gap of 1.29 eV
between the conduction and valence bands with respect to $\sigma$.
It can be seen that $\Delta_s$ decreases with increasing $|\sigma|$.
The electronic occupation numbers of the lower spin-up and the
higher spin-down quasilocal states were defined as $n^+$ and $n^-$,
respectively. It can be seen that under electron doping ($\sigma <
0.0$), the spin-up and spin-down quasilocal states are fully and
partially occupied ($n^+ = 1.0$, $n^- < 1.0$), respectively, while
under hole doping ($\sigma > 0.0$), they are partially and not
occupied ($n^+ < 1.0$, $n^- = 0.0$), respectively. The variations of
$\Delta_s$ and $\Delta{n}$ ($= n^+-n^-$) with respect to $\sigma$
are shown in Fig. \ref{DOS_model}. The curve of $\Delta_s(\sigma)$
is strictly linear and symmetric around $\sigma = 0.0$. A simple
exchange model can be used to express $\Delta_s$ as
\begin{eqnarray}{\label{exchange_model_1}}
\Delta_s=\epsilon^--\epsilon^+=-J_{eff}^-n_-S^-S^--(-J_{eff}^+)n_+S^+S^+
\end{eqnarray}
where $\pm$ represent spin-up and spin-down quasilocal states,
respectively; $\epsilon^\pm$ are the energies of the quasilocal
states; $J_{eff}^\pm$ are the exchange constants; $S^\pm =
\pm\frac{1}{2}$. In the exchange model, electrons with opposite
spins do not interact with each other. The linearity and symmetry of
the curve of $\Delta_s(\sigma)$ indicate that the exchange constant
does not vary with the spin orientation or $\sigma$, which can be
expressed as $J_{eff}^+(\sigma) = J_{eff}^-(\sigma) = J_{eff}$.
Then, Equ. \ref{exchange_model_1} can be rewritten as
\begin{eqnarray}{\label{exchange_model_2}}
\Delta_s=\frac{1}{4}J_{eff}{\Delta}n
\end{eqnarray}
The magnetic moment of the system equals $\Delta{n}$ $\mu_B$. The
curve $\Delta_s(\sigma)$ can be well fitted by Equ.
\ref{exchange_model_2} with $J_{eff}$ taken to be 1.23 eV, as shown
in Fig. \ref{DOS_model}.

\section{CONCLUSIONS}
\par The thermodynamic and kinetic properties of H and D
monomers on doped graphene layers were studied using a composite
method consisting of density functional theory, density functional
perturbation theory and harmonic transition state theory. The
calculated results were analyzed with reference to the electronic
structures and phonon spectra. Electron doping has been found to
heighten the diffusion potential barrier, while hole doping lowers
it. However, both kinds of dopings heighten the desorption potential
barrier. These phenomena in energies have been found to be
conjunctly due to the effects of charge doping on the strength of
the $\Pi$ bond (defined to include the $\pi$ and $\pi^*$ bonds) in
graphene and to the electron transfer and the coulomb interaction
between the hydrogen monomer and graphene. It has been found that
hole doping is necessary for the observation of the diffusion of H
and D monomers on graphene. The kinetic properties of H and D
monomers on doped graphene layers during both the annealing process
($t_0 = 300$ s) and the constant-rate heating process ($\alpha =$1.0
K/s) were simulated. H monomer is more mobile than D monomer in the
kinetic simulations. Generally, both electron doping and hole doping
can increase the desorption temperatures of hydrogen monomers.
However, the diffusion of hydrogen monomers is prevented by the
electron doping and triggered by the hole doping, and the diffusion
radius increases with $\sigma$ under hole doping. It has been found
that the macroscopic diffusion of hydrogen monomers can be achieved
at temperatures below the desorption temperature when $\sigma$
reaches $5.0\times10^{13}$ cm$^{-2}$ in graphene. The effect of
charge doping on the magnetic properties of the hydrogenated
graphene were also studied. The exchange splitting of the spin-up
and spin-down quasilocal states and the magnetic moment decrease
linearly with the number of doped electrons (holes). The variation
of the exchange splitting with respect to $\sigma$ has been
explained by a simple exchange model, where the exchange constant
has been found not to vary with the spin orientation or the doping
charge density.

\begin{acknowledgments}
The first author (Huang) wishes to thank Liv Horkek{\ae}r for
helpful email exchanges. This work was supported by the special
Funds for Major State Basic Research Projects of China (973) under
grant No. 2007CB925004, 863 Project, Knowledge Innovation Program of
Chinese Academy of Sciences and by Director Grants of CASHIPS. Part
of the calculations were performed at the Center of Computational
Science of CASHIPS and at the Shanghai Supercomputer Center.
\end{acknowledgments}

\bibliography{basename of .bib file}


\end{document}